\documentclass{article}
\usepackage{graphicx} 
\usepackage{natbib}
\usepackage{amsmath}
\title{Endogenous Crashes as Phase Transitions}
\author{
{\large Revant Nayar}$^{a,}$\footnote{Corresponding author at: FMI Technologies LLC, New York, United States. E-mail address: \texttt{rnayar@fmitech.net}.} \hspace*{2mm}
{\large Minhajul Islam}$^{a}$
\\[3mm]
$^{a}$\textit{\small FMI Technologies LLC, New York, United States}}

\date{August 2024}

\begin{document}

\maketitle
\begin{abstract}
    This paper explores the mechanisms behind extreme financial events, specifically market crashes, by employing the theoretical framework of phase transitions. We focus on endogenous crashes, driven by internal market dynamics, and model these events as first-order phase transitions—critical, stochastic, and dynamic. Through a comparative analysis of early warning signals associated with each type of transition, we demonstrate that dynamic phase transitions (DPT) offer a more accurate representation of market crashes than critical (CPT) or stochastic phase transitions (SPT). Unlike existing models, such as the Log-Periodic Power Law (LPPL) model, which often suffers from overfitting and false positives, our approach grounded in DPT provides a more robust prediction framework. Empirical findings, based on an analysis of S\&P 500 stocks from 2019 to 2024, reveal significant trends in volatility and anomalous dimensions before crashes, supporting the superiority of the DPT model. This work contributes to a deeper understanding of the predictive signals preceding market crashes and offers a novel perspective on their underlying dynamics.
   
\end{abstract}
\section{Introduction}
Extreme events, such as market crashes, are pivotal moments in financial markets that can lead to widespread economic consequences. Understanding and predicting these events has long been a challenge for economists and financial theorists. Traditional models often view crashes as stochastic anomalies or as outcomes driven by external shocks. However, an emerging perspective suggests that many crashes are endogenous, arising from the internal dynamics of the market itself.

In this paper, we propose a novel approach to analyzing market crashes by framing them as phase transitions, a concept rooted in statistical physics. Phase transitions occur when a system undergoes a fundamental change in state, often triggered by small perturbations that lead to large-scale, systemic shifts. In the context of financial markets, we hypothesize that crashes can be seen as first-order phase transitions, categorized into three types: critical, stochastic, and dynamic. Each type corresponds to different underlying mechanisms and early warning signals.

Critical phase transitions (CPT) have been extensively studied in various fields, including ecology and economics, where they are characterized by a slowing down of system responses as a critical threshold is approached. Stochastic phase transitions (SPT), on the other hand, occur when randomness or noise within the system amplifies fluctuations, leading to sudden shifts. Dynamic phase transitions (DPT), less explored in the financial context, involve changes in the underlying noise distribution itself, potentially driven by feedback loops within the market.

Our analysis compares these three types of phase transitions in the context of financial crashes. We assess their early warning signals and evaluate their predictive power using empirical data from the S\&P 500. We also compare our findings with the widely studied Log-Periodic Power Law (LPPL) model, which has been used to predict crashes but suffers from limitations such as overfitting and a high rate of false positives.

The results of our study suggest that dynamic phase transitions provide a more accurate and generalizable framework for understanding market crashes. By examining trends in volatility, skewness, and anomalous dimensions, we demonstrate that the signals associated with DPT are more consistent with the observed behavior of markets leading up to crashes. This work not only advances the theoretical understanding of market crashes but also offers practical tools for early detection and risk management.

In the following sections, we will delve deeper into the theoretical foundations of phase transitions, describe our empirical methodology, and discuss the implications of our findings for both financial theory and practice.

\section{Endogenous Versus Exogenous Crashes}
 Usually, market crashes or extreme events in general belong to two dynamical classes. Endogenous crashes are caused by the internal dynamics of the system (in the case of markets, the cycles of demand and supply); exogeneous crashes are caused by external news events. These belong to different dynamical classes; this can be seen from volatility and volume dynamics. We will deal solely with endogeneous; specifically bubble-induced shocks and crashes. We can have exogenous shocks (eg. COVID-related in March 2020), which this does not apply to, because naturally, they have no Early Warning Signals. In a bubble, the observed price trajectory deviates from its intrinsic fundamental value. The market is irrationally exuberant- driven by sentiment, and the price no longer reflects underlying value. When liquidity starts to dry, banks raise rates, capital inflows stop, there is panic synchronised selling, causing the bubble to pop. The market becomes unstable like a ruler balanced on the tip of a finger and small shocks will cause the crash. This is a symptom of criticality, which is accompanied by an enhancement of symmetry

\section{Log-Periodic Power Law (LPPL) Model}
We begin with the assumption of a stochastic differential equation (SDE) where \( p \) represents the price:
\[
\frac{dp}{p} = \mu(t)dt + \sigma(t)dW - \kappa dj
\]
Here, \( dW \) is an infinitesimal increment of a Wiener process, and \( dj \) represents a discontinuous jump, where \( j = 0 \) before a crash and \( j = 1 \) after the crash. The parameter \( \kappa \) quantifies the amplitude of the crash when it occurs. Notably, \( \kappa \) can be negative, signifying an anti-bubble that leads to an anti-crash; this symmetry implies that the situation is fundamentally symmetric between crashes and anti-crashes.
The hazard rate \( h(t) \), which quantifies the probability of a crash, is defined as:
\[
\langle dj \rangle = h(t)dt
\]
Due to Discrete Scale Invariance (DSI), the hazard rate \( h(t) \) follows a power-law behavior with a complex exponent:
\[
h(t) \propto |t - t_c|^{\alpha + i \beta}
\]
This results in anomalous dimensions that have both real and imaginary components. By taking the real component, we obtain:
\[
h(t) = \alpha (t_c - t)^{m-1} \left( 1 + \beta \cos\left(\omega \ln(t_c - t) - \phi\right)\right)
\]
where the parameters are \( \alpha \), \( \beta \), \( \omega \), \( \phi \), and \( t_c \), the critical time at which the crash occurs.
Integrating this expression into the expectation value of the price change, we derive:
\[
\ln(p(t)) = A + B(t_c - t)^m + C(t - t_c)^m \cos\left(\omega \ln(t_c - t) - \phi\right)
\]
It is important to note that this formula is valid only up to the critical time \( t_c \) and not beyond.
This model can be fitted by minimizing the least squares residual:
\begin{align*}
S(t_c, m, \omega, \phi, A, B, C) &= \sum \left[\ln(p(t)) - \left(A + B(t_c - t)^m \right. \right.\\
&\quad \left. \left. + C(t - t_c)^m \cos\left(\omega \ln(t_c - t) - \phi\right)\right)\right]^2
\end{align*}
This equation involves three linear and four nonlinear parameters. The optimization process can be reformulated to involve three nonlinear and four linear parameters:
\begin{align*}
S(t_c, m, \omega, \phi, A, B, C) &= \sum \left[\ln(p(t)) - \left(A + B(t_c - t)^m \right. \right.\\
&\quad \left. \left. + C_1 (t - t_c)^m \cos\left(\omega \ln(t_c - t)\right) \right.\right. \\
&\quad \left. \left. + C_2 (t - t_c)^m \sin\left(\omega \ln(t_c - t)\right)\right)\right]^2
\end{align*}
where \( C_1 = C \sin(\phi) \) and \( C_2 = C \cos(\phi) \).
\begin{figure}[h]
  \centering
  \includegraphics[width=80mm]{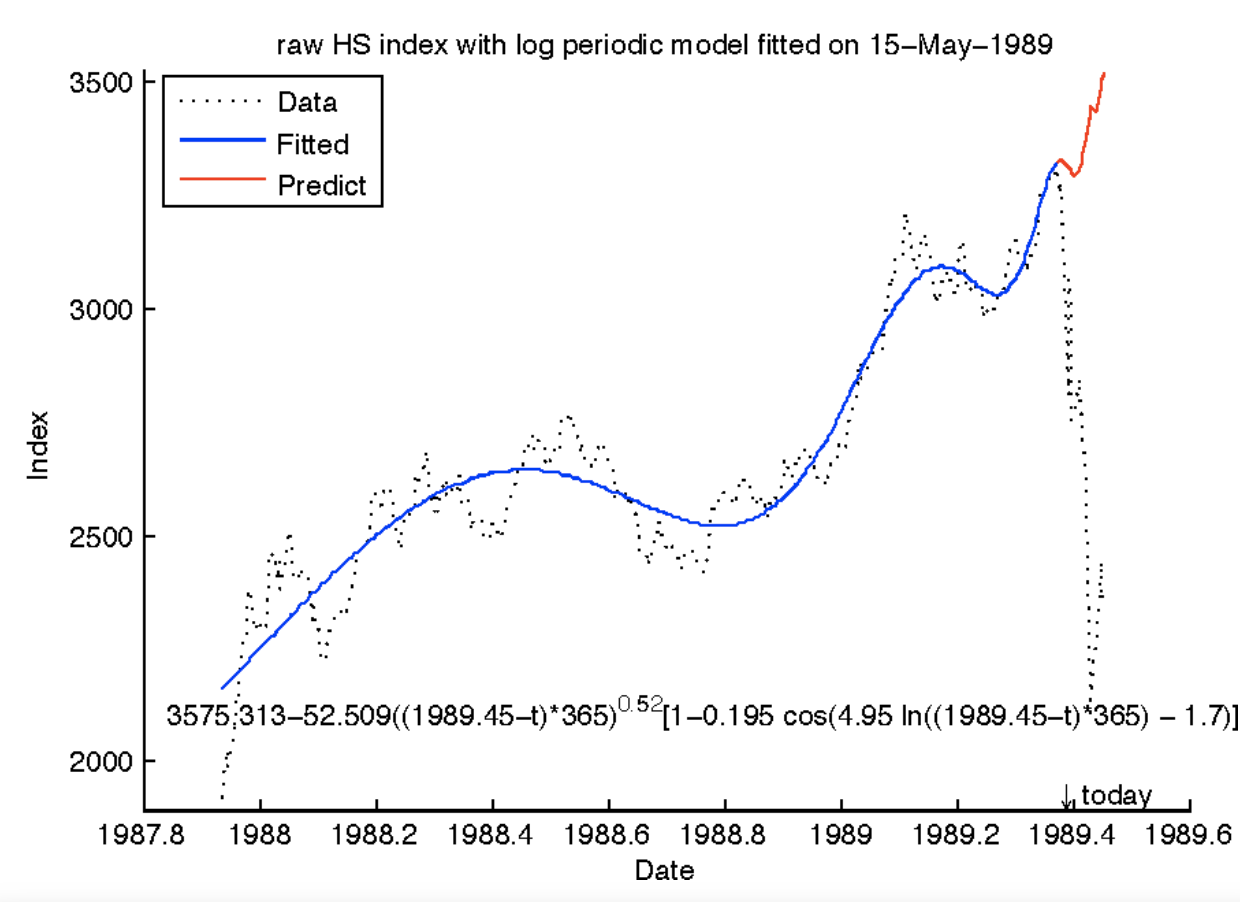}
  \caption{Log-Periodic Power Law Price Formation}
\end{figure}
There are several notable weaknesses in this model. First, the introduction of a jump by hand into the SDEs to explain endogenous crashes may be overly simplistic. Second, the model’s reliance on seven parameters increases the risk of overfitting. Indeed, some studies have found LPPL signatures even in independently and identically distributed (iid) noise data. These weaknesses can be addressed and refined, as discussed in subsequent sections.

\section{Phase transitions}
Let us start with a generic SDE:
$\\dp=-h(p(t))dt+\sigma dW_{t}\\$
There are three different components that we can examine- the drift $h$, the diffusion term $\sigma$ and the noise term $W_{t}$. 
This is motivated by Halperin's non-equilibrium skew (NES) model, as well as Guttal's models. Usually, qualitative results remain the same even if we have higher-order terms.
There are three ways to get a crash
\begin{enumerate}

\item Make the drift term time-dependent $h(p(t)) \to h(p(t),t)$, keeping the volatility and noise fixed (Critical Slowing Down)
         \item Make volatility time-dependent and increasing $\sigma \to \sigma(t)$, keeping the potential and noise fixed (Stochastic Phase Transition)
        \item Evolve the noise term, keeping the drift and volatility fixed $dW \to d\alpha$ (Dynamical Phase Transition; no SDE/Lagrangian description)
\end{enumerate}

\subsection{Critical Phase Transition}
Without losing generality we can expand the drift term up to cubic order:
$\\dp=(-\mu(t)+rp-p^{3})dt+\sigma dW_{t}\\$
Here we can assume different forms of the time dependence. In any case, we see:
\begin{itemize}
\item Increasing Volatility
\item Increasing Negative (or Positive) Skewness
\item Increasing autocorrelation at lag 1(critical slowing down)
\item Decreasing instantaneous frequency
\end{itemize}
\begin{figure}
  \centering
  \includegraphics[width=50mm]{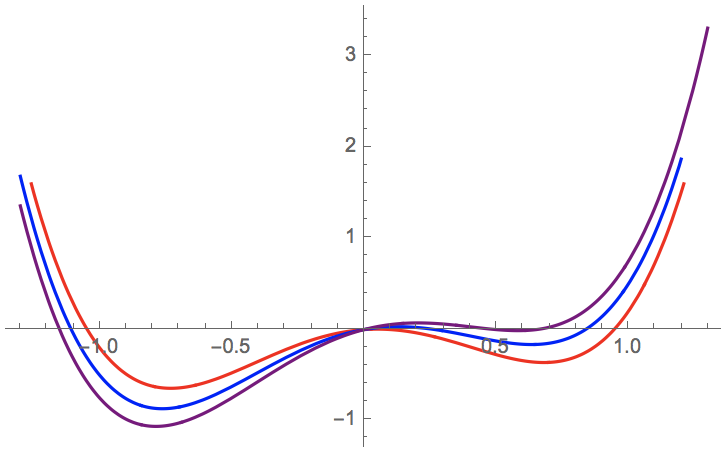}
  \caption{Change of Potential with time in critical phase transition}
 \end{figure}
\subsection{Stochastic Phase Transition}
We can have the volatility change with time:
$\\dp=-h(p,t) dt+\sigma(t) dW_{t}\\$
Let's say we postulate a linear increase:
$\\\sigma(t)=\alpha t, t<t_{c}\\$
As the volatility increases, we can increase, we increase the probability that we tunnel from one potential minimum to the other. We see the following early warning signals:
\begin{itemize}
\item 1. Increasing Volatility
\item 2. Increasing (Positive or Negative) Skewness
\end{itemize}
\begin{figure}
  \centering
  \includegraphics[width=50mm]{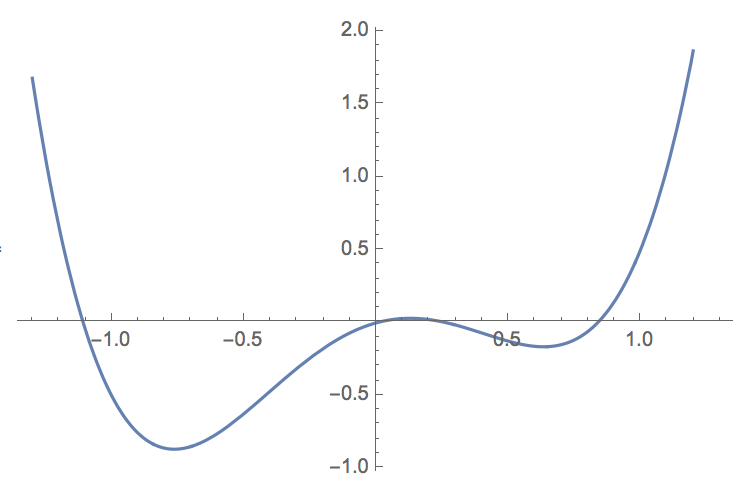}
  \caption{Fixed Potential in stochastic phase transition}
 \end{figure}

\section{Dynamic Phase Transition}
In this framework, the noise term itself evolves over time, becoming increasingly fat-tailed. This behavior is akin to an $\alpha$-stable distribution where $\alpha(t)$ transitions from 2 to 1 during times of crashes. Alternatively, it can be modeled as a fractional Brownian motion (fBM) with the Hurst exponent transitioning from $H=0.5$ to $H=1$. These dynamics are not easily described by any known partial differential equation (PDE) or stochastic differential equation (SDE). However, this approach is motivated by a simple observation: during crises, time and log-price ($x$) scale uniformly. Price dynamics become invariant under the transformation
\[
x \to \lambda x, \quad t \to \lambda^{z} t,
\]
where a crash corresponds to $z$ transitioning from 2 to 1. This scaling gives rise to the following observables.
\subsection{Observables}
For self-similar processes, we can constrain the form of expectation values $\langle \cdot \rangle$. Assuming self-similarity (conformal symmetry), we have
\[
\langle X(t) \rangle = |t|^{\Delta}, \quad \langle X(t)X(t+\tau) \rangle = \frac{c}{|\tau|^{2 \Delta}}.
\]
In this context, a crash is represented by $\Delta = \frac{1}{2} \to 1$. The challenge now lies in measuring the time dynamics of $\Delta = \Delta(t,\tau)$. We can express this as:
\[
\ln\langle X(t)X(t+\tau) \rangle = K - \Delta(t,\tau) \ln(|\tau|),
\]
from which we derive:
\[
\Delta(t,\tau) = \frac{K - \ln(G(t,\tau))}{\ln \tau}.
\]
We allow for multifractality, which corresponds to non-trivial Generalized Hurst Exponents (GHE), or 'Hursts' of higher-order operators:
\[
\langle X(t)^{n} \rangle = A_{n} |t|^{\Delta_{n}}, \quad \text{where} \quad \Delta_{n} = \frac{K_{n} + \ln\langle X(t)^{n} \rangle}{\ln(|t|)}.
\]
In the multivariate case, we have:
\[
\langle X^{m}(t)X^{n}(t+\tau) \rangle = \frac{c_{mn}}{|\tau|^{\Delta_{m} + \Delta_{n}}}.
\]
Changing variables as before:
\[
\Delta_{m} + \Delta_{n} = \frac{K - \ln\langle X^{m}(t)X^{n}(t+\tau) \rangle}{\ln \tau}.
\]
We observe nonlinearity (multifractality) if $\Delta_{n} \neq n \Delta$, and non-stationarity if $\Delta_{k} \to \Delta_{k}(t)$. Additionally, we can break conformality by introducing $\tau$ dependence: $\Delta_{k}(t) \to \Delta_{k}(t,\tau)$. We define a conformality index that captures how much $\Delta(t,\tau)$ varies with $\tau$.

In this study, however, we break stationarity while preserving conformality. We conduct empirical analyses across a range of crashes and derive trends in anomalous dimensions before crashes versus at other times. We find a significant increase in conformality before a crash, in contrast to a general decrease not associated with crashes. This approach does not require a model of PDE or SDE; it is completely independent of such models, which is a distinct advantage of methods derived from conformal field theory, as they tend to be more general.

In addition to an increase in the anomalous dimension, we also observe an increase in volatility, where $\sigma \propto |t|^{\Delta_{2}}$ and $\Delta_{2}$ increases with time, leading to a corresponding increase in $\sigma$.

\subsection{Empirical Findings}
Here is a summary of empirical findings based on S\&P 500 stocks over five years between January 2019 and January 2024. Here a crash is defined as a 20 percent drop in the price of any stock. We plot trends in the early warning signals averaged before the crashes, and averaged at other times. Our findings are as follows:
\begin{itemize}
    \item A weak positive trend is seen in volatility before crashes

    \item A weak negative trend is seen in skewness before crashes

    \item No trend is seen in lag 1 autocorrelation, or critical slowing down.

    \item A strong trend is seen in the anomalous dimension 
\end{itemize}
The findings support a case for dynamic, as opposed to critical or stochastic phase transitions. 
\begin{figure}
  \centering
  \includegraphics[width=50mm]{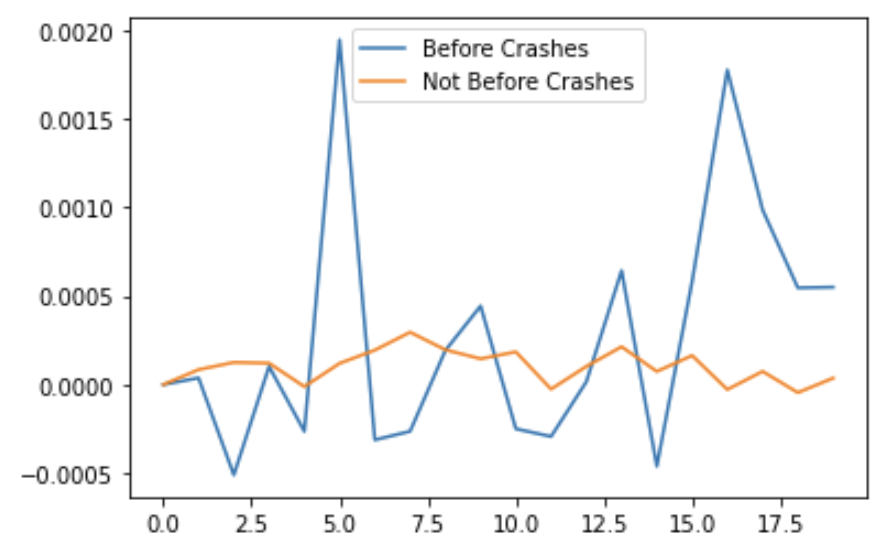}
  \caption{Trend in Volatility before Crashes (Versus normal times)}
 \end{figure}
 \begin{figure}
  \centering
  \includegraphics[width=50mm]{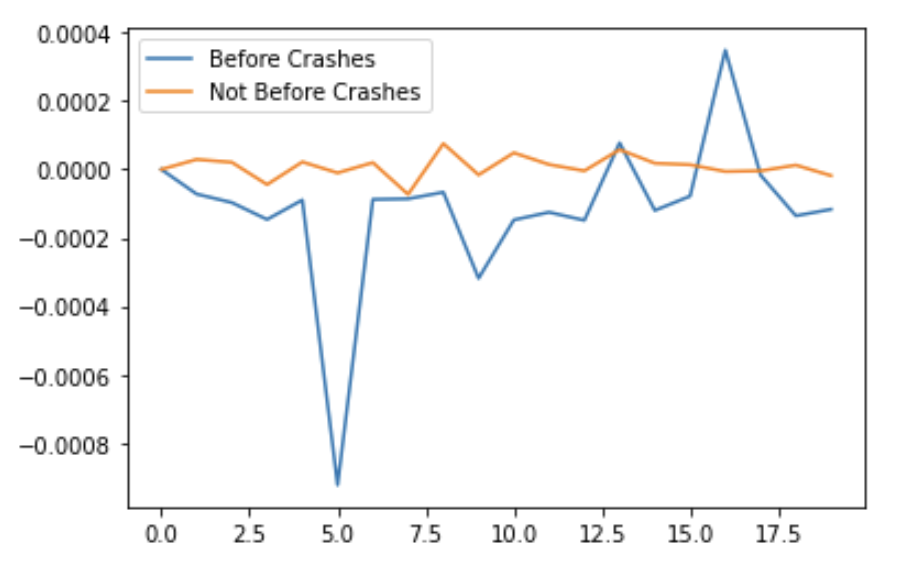}
 \caption{Trend in Skewness before Crashes (Versus normal times)}
 \end{figure}
 \begin{figure}
  \centering
  \includegraphics[width=50mm]{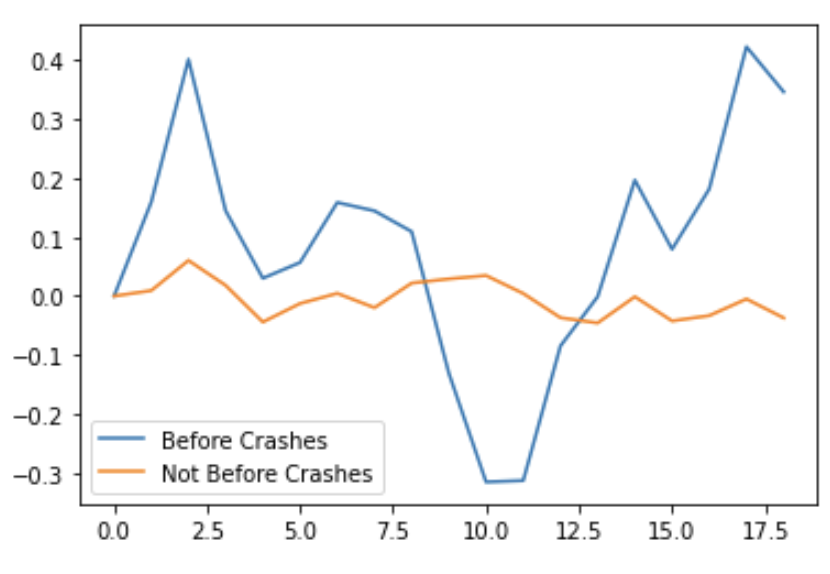}
  \caption{Trend in Lag 1 Autocorrelation before Crashes (Versus normal times)}
 \end{figure}
 \begin{figure}
  \centering
  \includegraphics[width=50mm]{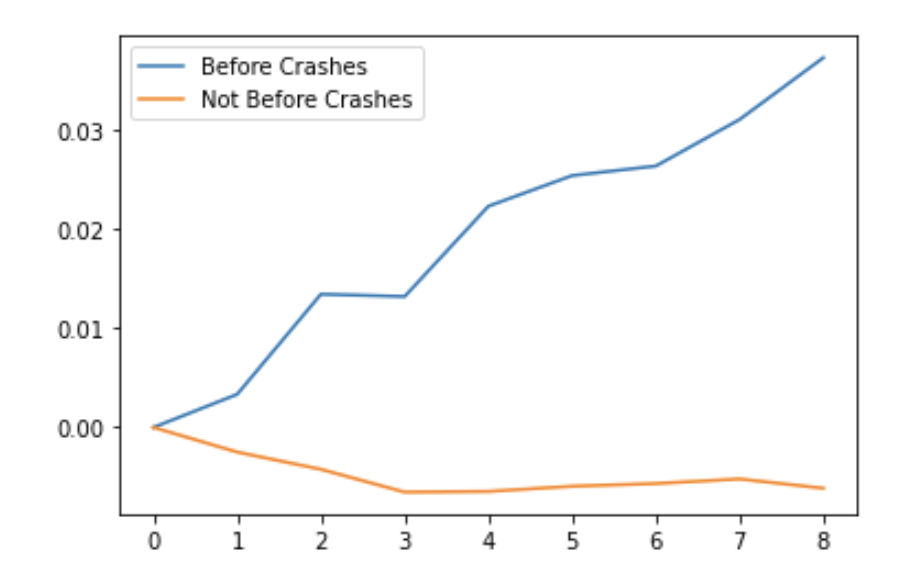}
  \caption{Trend in Anomalous Dimensions before Crashes (Versus normal times)}
 \end{figure}

\section{Multivariate Case}
We expect to observe signals of a crash by examining early warning signals (EWS) at the level of financial indices, which are weighted averages of the markets and sectors. It has also been noted that there is an increase in cross-correlations before market crashes, as analyzed using network theory and the eigenvalues of the correlation matrix. However, these observations are usually not derived from first principles. We can model the multivariate case as a series of stochastic differential equations (SDEs):
\[
dp_{i} = V_{i}(p_{i}) \, dt + \sigma_{i} \, dW_{i},
\]
where
\[
V_{i}(p_{i}) = -\mu(t) + r_{i}p_{i} - \lambda_{i} p_{i}^{3}.
\]
These equations are coupled to each other through the noise term:
\[
\langle dW_{i} dW_{j} \rangle = D_{ij},
\]
where \(D_{ij}\) captures the connectivity properties of the system and is almost always positive for stocks, whereas \(\langle V_{i}(p_{i})V_{j}(p_{j}) \rangle\) can be either positive or negative. 
As the system approaches critical slowing down in the form of a first-order phase transition, we find that \(V_{i}(p_{i}) \approx 0\) and \(\langle V_{i}(p_{i})V_{j}(p_{j}) \rangle \approx 0\). Therefore, 
\[
\langle dp_{i} dp_{j} \rangle \approx D_{ij}.
\]
This implies that \(\langle dp_{i} dp_{j} \rangle\) generally increases with time as the system approaches criticality.
Empirical studies, particularly in ecological systems, have shown that for highly connected systems, spatial correlation is a stronger early warning signal than temporal autocorrelation. This observation is consistent with the behavior of systems undergoing critical phase transitions. Moreover, the covariance between the Brownian motions is given by:
\[
\text{Cov}(W^{H_{1}}, W^{H_{2}}) = |t|^{H_{1}+H_{2}}.
\]
As the Hurst exponents \(H_{i}\) increase during a dynamic phase transition, cross-covariances also increase. Therefore, the observed increase in cross-covariance is expected under both critical and dynamic phase transitions. However, such an increase is not expected under stochastic phase transitions or within the Log-Periodic Power Law (LPPL) framework.
\begin{figure}[h]
  \centering
  \includegraphics[width=50mm]{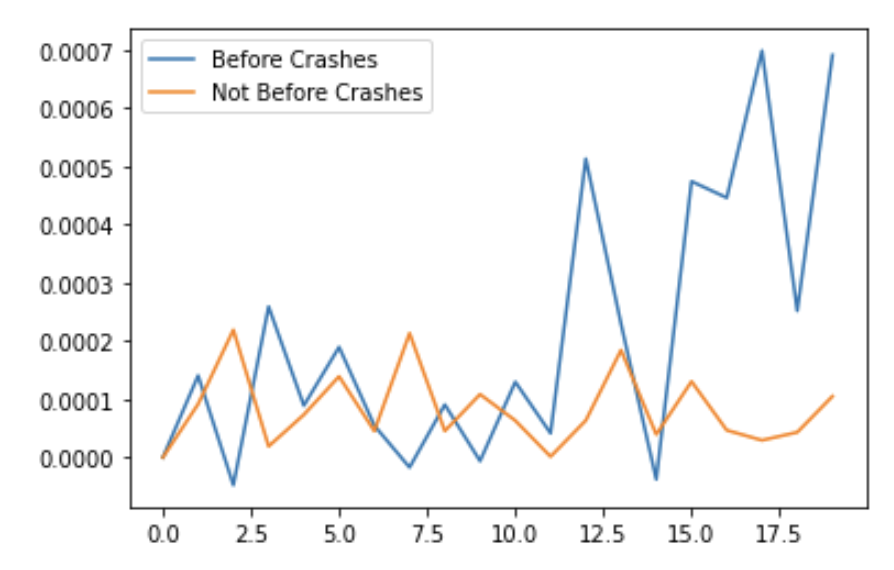}
  \caption{Trend in Cross Covariance before Crashes (versus normal times)}
\end{figure}

\section{Conclusion}

This paper has introduced a novel framework for understanding financial market crashes by modeling them as first-order phase transitions. By examining the internal dynamics of markets, we have classified crashes into three distinct types: critical, stochastic, and dynamic phase transitions. Our comparative analysis has demonstrated that dynamic phase transitions (DPT) offer a more robust and accurate model for predicting market crashes compared to critical (CPT) and stochastic phase transitions (SPT), as well as the widely used Log-Periodic Power Law (LPPL) model.

Empirical analysis using S\&P 500 data over a five-year period revealed significant early warning signals associated with dynamic phase transitions, particularly in the form of increasing volatility and changes in anomalous dimensions. These findings suggest that market crashes are not merely random or externally driven events but are often the result of endogenous processes that can be anticipated through careful analysis of market indicators.

The implications of this study are twofold. First, it advances the theoretical understanding of market crashes by integrating concepts from statistical physics into financial modeling. Second, it provides practical tools for risk management and early detection, potentially allowing market participants to better prepare for and mitigate the impact of crashes.

Future research could extend this framework by exploring its applicability to other types of financial markets and instruments, as well as by refining the models to account for additional variables or more complex market dynamics. Additionally, further empirical validation across different time periods and market conditions would help solidify the practical relevance of these findings.

In conclusion, by framing market crashes as phase transitions, we offer a new lens through which to view and anticipate these critical events, paving the way for more resilient financial systems and better-informed market strategies.
 
\nocite{*}
\bibliographystyle{plainnat}  
\bibliography{name}  


\end{document}